\documentclass[sigconf]{acmart}

\AtBeginDocument{%
  }

\usepackage{multirow}%
\usepackage{graphicx}%
\usepackage{amsmath}
\usepackage{subfigure}
\usepackage{subcaption}
\setcopyright{none}
\renewcommand\footnotetextcopyrightpermission[1]{}
\begin{document}
%%
%% The "title" command has an optional parameter,
%% allowing the author to define a "short title" to be used in page headers.
\title{CATS: Clustering-Aggregated and Time Series for Business Customer Purchase Intention Prediction}
%%
%% The "author" command and its associated commands are used to define
%% the authors and their affiliations.
%% Of note is the shared affiliation of the first two authors, and the
%% "authornote" and "authornotemark" commands
%% used to denote shared contribution to the research.
%\author{Ben Trovato}
%\authornote{Both authors contributed equally to this research.}
%\email{trovato@corporation.com}
%\orcid{1234-5678-9012}
%\author{G.K.M. Tobin}
%\authornotemark[1]
%\email{webmaster@marysville-ohio.com}
%\affiliation{%
%  \institution{Institute for Clarity in Documentation}
%  \city{Dublin}
%  \state{Ohio}
%  \country{USA}
%}

\author{Yingjie Kuang}
\affiliation{%
  \institution{South China Agricultural University}
  \city{Guangzhou}
  \country{China}}
\email{kuangyj@scau.edu.cn}
\orcid{0009-0001-9115-4106}

\author{Tianchen Zhang}
\affiliation{%
  \institution{South China Agricultural University}
  \city{Guangzhou}
  \country{China}}
\email{ztc666@stu.scau.edu.cn}
\orcid{0009-0001-8440-4488}

\author{Zhen-Wei Huang}
\affiliation{%
  \institution{South China Agricultural University}
  \city{Guangzhou}
  \country{China}}
\email{zhenwhuang@gmail.com}
\orcid{0009-0009-2941-3152}

\author{Zhongjie Zeng}
\affiliation{%
  \institution{Wens Foodstuff Group Co., Ltd.}
  \city{Yunfu, Guangdong}
  \country{China}}
\email{zengzhjgdut@gmail.com}
\orcid{0009-0002-3339-1004}

\author{Zhe-Yuan Li}
\affiliation{%
  \institution{South China Agricultural University}
  \city{Guangzhou}
  \country{China}}
\email{lizheyuan@stu.scau.edu.cn}
\orcid{0009-0006-6813-3637}

\author{Ling Huang}
\affiliation{%
  \institution{South China Agricultural University}
  \city{Guangzhou}
  \country{China}}
\email{huanglinghl@hotmail.com}
\orcid{0000-0001-5089-4637}

\author{Yuefang Gao}
\authornote{Corresponding authors.}
\affiliation{%
  \institution{South China Agricultural University}
  \city{Guangzhou}
  \country{China}}
\email{gaoyuefang@scau.edu.cn}
\orcid{0000-0003-4794-9961}

\renewcommand{\shortauthors}{Yingjie Kuang, et al.}
%%
%% By default, the full list of authors will be used in the page
%% headers. Often, this list is too long, and will overlap
%% other information printed in the page headers. This command allows
%% the author to define a more concise list
%% of authors' names for this purpose.
% \renewcommand{\shortauthors}{Anonymous Author, et al.}

%%
%% The abstract is a short summary of the work to be presented in the
%% article.
\begin{abstract}
 Accurately predicting customers' purchase intentions is critical to the success of a business strategy. Current researches mainly focus on analyzing the specific types of products that customers are likely to purchase in the future, little attention has been paid to the critical factor of whether customers will engage in repurchase behavior. Predicting whether a customer will make the next purchase is a classic time series forecasting task. However, in real-world purchasing behavior, customer groups typically exhibit imbalance - i.e., there are a large number of occasional buyers and a small number of loyal customers. This head-to-tail distribution makes traditional time series forecasting methods face certain limitations when dealing with such problems. To address the above challenges, this paper proposes a unified Clustering and Attention mechanism GRU model (CAGRU) that leverages multi-modal data for customer purchase intention prediction. The framework first performs customer profiling with respect to the customer characteristics and clusters the customers to delineate the different customer clusters that contain similar features. Then, the time series features of different customer clusters are extracted by GRU neural network and an attention mechanism is introduced to capture the significance of sequence locations. Furthermore, to mitigate the head-to-tail distribution of customer segments, we train the model separately for each customer segment, to adapt and capture more accurately the differences in behavioral characteristics between different customer segments, as well as the similar characteristics of the customers within the same customer segment. We constructed four datasets and conducted extensive experiments to demonstrate the superiority of the proposed CAGRU approach.
\end{abstract}

%\begin{CCSXML}
%<ccs2012>
%   <concept>
%       <concept_id>10010405.10003550</concept_id>
 %      <concept_desc>Applied computing~Electronic commerce</concept_desc>
 %      <concept_significance>500</concept_significance>
%       </concept>
% </ccs2012>
%\end{CCSXML}

%\ccsdesc[500]{Applied computing~Electronic commerce}

% \section{CCS Concepts and User-Defined Keywords}

% Two elements of the ``acmart'' document class provide powerful
% taxonomic tools for you to help readers find your work in an online
% search.

%%
%% The code below is generated by the tool at http://dl.acm.org/ccs.cfm.
%% Please copy and paste the code instead of the example below.
%%

%%
%% Keywords. The author(s) should pick words that accurately describe
%% the work being presented. Separate the keywords with commas.
%% A "teaser" image appears between the author and affiliation
%% information and the body of the document, and typically spans the
%% page.

%%
%% This command processes the author and affiliation and title
%% information and builds the first part of the formatted document.
\maketitle

\section{Introduction}
\label{sec:introduction}
\begin{figure}[!t]
	\centering
	\centerline{
		\includegraphics[width=1.05\linewidth]{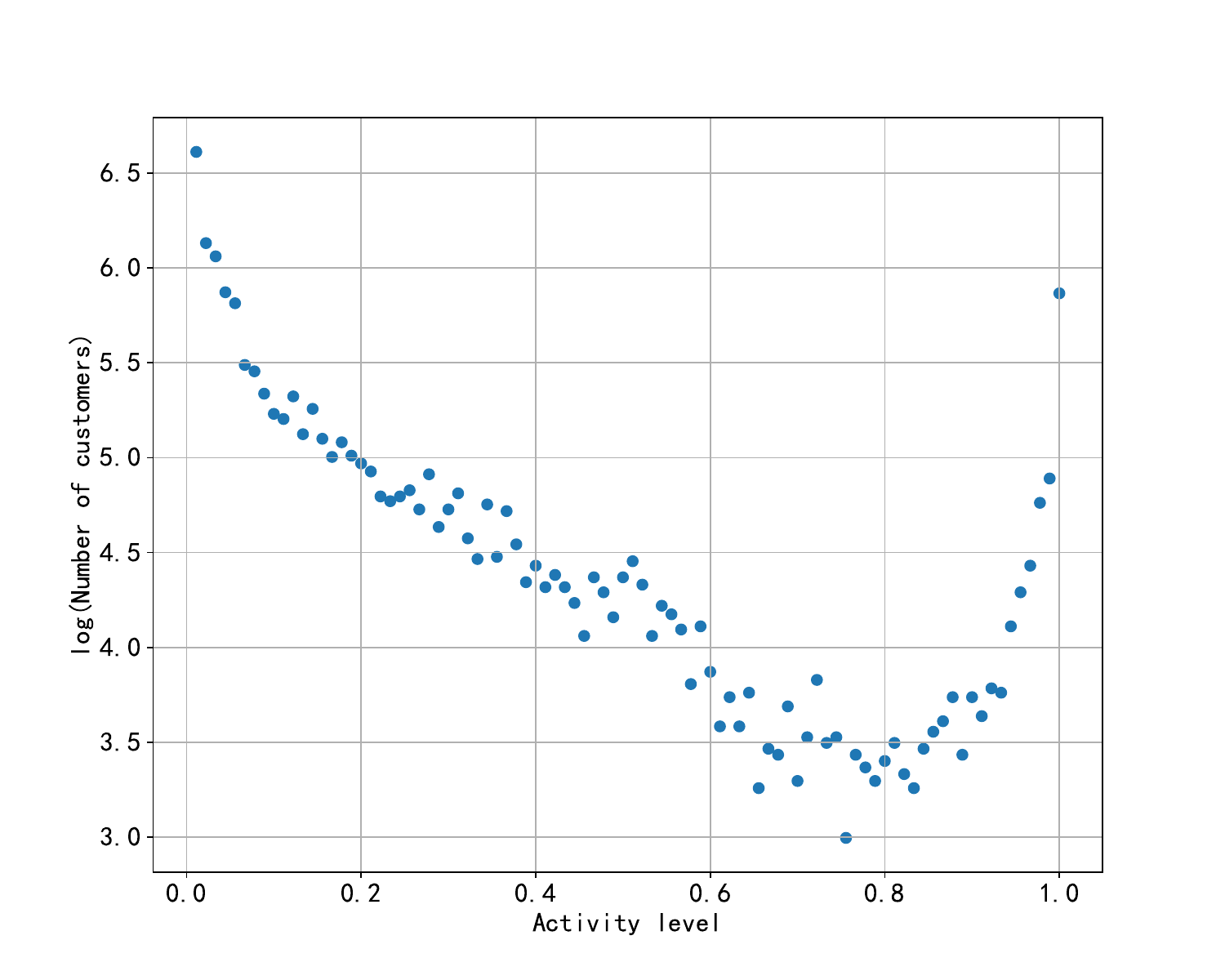}
	}
	\caption{Distribution of the relationship between customer activity and customer quantity. }
	\label{fig:CustActiveness}
\end{figure}

Customer purchase intention prediction is a time-series prediction task based on historical customer behavior data to determine whether a customer will make a purchase in a future period~\cite{luo2019user}. In a realistic model of e-commerce, the groups of customers will tend to be very diverse. As illustrated in \figurename~\ref{fig:CustActiveness}, the distribution of the number of customers corresponding to different activity levels in the past 100 days of an e-commerce enterprise is counted. It contains customer groups with different degrees of activity such as low, medium and high, and generally shows the phenomenon of head-to-tail distribution. Most of the customer population are mainly concentrated in the low and high activity zones, which correspond to occasional customers and loyal customers, respectively. For enterprises, on the one hand, they need to prevent the loss of loyal customers. On the other hand, it is necessary to dig out potential effective customers in the customer base. In the medium and low active customers there may be a low frequency of purchase but the transaction amount is large, and there is a regularity of the purchase pattern of the potential value of the customer. Therefore, the ability of effectively predicting the purchase intention of customers is very important for enterprises to make decision management. However, customers' purchase intentions are subjectively influenced by factors such as interest drive~\cite{srivastava2023customer}, satisfaction~\cite{artana2022repurchase}, and loyalty~\cite{meyer2008influence}, and they are objectively susceptible to external factors such as market activities and seasonal changes. These complexities make accurate modeling of customers a huge challenge.

Based on different processing strategies, existing approaches to customer purchase intention can be broadly summarized as statistical models, machine learning models, and deep learning models. Statistical methods make predictions by analyzing in advance the relationship between the influencing factors and the prediction objectives~\cite{bringula2018factors, bag2019predicting}. However, it is difficult for a single statistical model to achieve the desired prediction accuracy because the complex relationships between variables are often difficult to describe in advance. The machine learning methods can alleviate the overfitting problem that statistical models are prone to~\cite{saranya2020prediction, esmeli2021towards, khandokar2023gradient}. However, the methods cannot achieve positive results when facing data imbalance problems such as the intertemporal and highly sparse data distributions that occur in large-scale datasets. Moreover, it is time-consuming and difficult to capture customer behavioral preferences. Owing to the time effect of customer behavior, methods based on deep learning models have been widely used in recent years~\cite{liu2021tpgn, huang2021purchase, chaudhuri2021platform, khattak2021applying}. Deep learning methods effectively capture the long and short-term preferences of customers and alleviate the data distribution imbalance problem, achieving better customer prediction results.

Although these methods have achieved good results, they focus on the characteristics of the customers themselves or among them and do not take into account the differences between groups of customers. In terms of purchasing behavior, loyal customers usually show strong repeat purchasing behavior, which is usually influenced by the demand for a stable supply of certain products over time. And there are potential groups among the medium and low active people whose buying patterns are characterized by certain regularity. It is a challenge to accurately identify customer groups and the purchasing patterns of different customer segments. Potential connections between customers and differences between customer groups need to be considered in customer modeling. In order to cope with these challenges, this paper proposes a new approach to customer model based on clustering for the phenomenon of head-to-tail distribution of customer groups as well as the regularity, similarity, and other characteristics of customers' purchasing behaviors.

In this paper, we analyze customer behavior, demonstrating the feasibility of clustering customers to improve the prediction accuracy. For the prediction of customers' purchase intentions, we propose a unified Clustering and Attention mechanism GRU model (CAGRU). Specifically, the method first constitutes a time series of the customer's historical purchase records, and the node of each sequence is represented as the corresponding supplier of the order placed by the customer on the same day. Through the company sales records and multi-modal data, a multivariate time series can be constructed between the customers to mine out their similarities. Subsequently, the customers are clustered using the k-shape clustering algorithm, which delineates distinct clusters of customers exhibiting similar characteristics. Then, a strategy combining attention and GRU is used to extract time series features and predict customers' purchase intentions, where the attention mechanism improves the prediction performance by capturing customers' short-term purchase preferences. To evaluate the effectiveness of the proposed model, we constructed four datasets with different customer sizes based on marketing data provided by a poultry company, and extensive experiments and ablation studies were conducted to demonstrate the effectiveness of our proposed approach. 

This work makes the following main contributions:
\begin{itemize}
\item In this paper, a customer survey of companies was launched to analyze the similarities and differences in purchasing behavior through clustering method, leveraging multi-modal data. 
\item With a focus on the phenomenon of head-to-tail distribution of customer groups and the characteristics of customers with multiple similarities and time effects, we propose a unified clustering and attention GRU model to achieve accurate prediction of customers' purchase intentions. 
\item We construct four poultry sales datasets with different customer sizes and conduct extensive comparisons with several state-of-the-art models to validate the superiority of the proposed model.
\end{itemize}

\section{Related Work}
\label{sec:RelatedWork}
\subsection{Customer Purchase Intention Prediction}
\label{sec:Customer Purchase Intention Prediction}
Customer purchase intention prediction is a key issue in marketing and customer relationship management in e-commerce. In recent years, researchers have proposed a variety of methods, such as decision trees, random forest, and gradient boosting trees~\cite{liu2020machine, yang2021rf, satu2023modeling}, which are mainly used to improve the prediction accuracy by capturing complex feature relationships and processing non-linear patterns. However, these models lack the ability to capture temporal features. To further enhance performance, many deep learning methods have been proposed~\cite{liu2021tpgn, huang2021purchase, ling2019customer, kim2021deep}. Recent work~\cite{huang2021purchase} developed a convolutional hierarchical transformer network to enable the modeling of buying patterns with multigranularity temporal dynamics. Different from the above methods, this paper uses clustering methods to capture the preferences of the different customer groups by analyzing the customers' multivariate similarities.

\subsection{Time Series Clustering and Forecasting}	
\label{sec:Time Series Clustering and Forecasting}
Time series clustering and forecasting play important roles in data analysis. The purpose of time series clustering is to group time series with similar patterns. This mainly includes the distance-based methods~\cite{petitjean2011global, paparrizos2015k, lloyd1982least}, feature-based methods~\cite{zhou2018clustering, du2018multivariate, liang2017clustering}, and deep learning-based methods~\cite{vaquez2020preliminary, yang2024bidirectional, ienco2020deep}. The goal of time series forecasting is to predict future values via historical data, and existing time series forecasting models are mainly based on deep learning models, such as RNNs~\cite{hochreiter1997long, chung2014empirical}, CNNs~\cite{bai2018empirical, franceschi2019unsupervised}, eMLPs~\cite{ekambaram2023tsmixer, das2023long}, Transformers~\cite{zhang2022crossformer, zhou2021informer}, and so on. These models show strong capabilities in feature extraction and processing non-linear relationships. Some works have also combined the above techniques~\cite{bandara2020forecasting, bhaskaran2020future, zhu2018correlation}, and the accuracy and the reliability of time series prediction can be significantly improved through the rational use of clustering information. For example, Bhaskaran et al.~\cite{bhaskaran2020future} proposed an oil pipeline fault identification and prediction method. This work used k-means to cluster the pipeline pressure data, followed by the use of linear regression combined with time series prediction to predict the future failure rate. In this paper, we combine time series clustering and forecasting to predict customers' purchase intentions. 

\section{Customer Survey and Problem Formulation}
\label{sec:User_Survey}
In this section, we analysis the regularity of customer behavior and the similarity between customers based on a real marketing dataset provided by a poultry company. To measure customer behavior information, we define the following key concepts.

\textbf{Customer Activeness:} To quantify the level of customer purchases, we define an indicator metric on the number of recent purchases made by customer $u$ for a company $c$ as follows:
\begin{equation}
     P (u, c)= \frac {\sum\limits_{i=1}^t d(u, c, i)}{t} ,\quad d(u, c, i) \in \{0, 1\}
\end{equation}
where $d(u, c, i)$ denotes whether customer $u$ visited company $c$ on day $i$, $t$ represents the total time duration, and $p(u, c)$ indicates the engagement level of customer $u$ with company $c$. In addition, we examine the relationship between company engagement and customer count. As illustrated in \figurename~\ref{fig:CustActiveness}, the relationship between engagement and customer count follows a long-tail distribution. Most customers belong to the category of nonactive customers and are considered nontarget customers in business terms. Another segment of customers exhibits moderate to high engagement levels, representing the target audience for businesses. These customers often demonstrate certain regularities and similarities in their attendance behavior. 

To further explore these properties, based on the historical purchase information of these customers, we select one of the companies and extract the purchase time series of the customers under the company, as shown below:
\begin{equation}
	X_u = \{d_u^1, \dots, d_u^t \},\quad  d_u^t \in \{0, 1\}
\end{equation}
where $X_u$ denotes the attendance behavior sequence of customer $u$ over $t$ days, and $d_u^t$ indicates whether customer $u$ visited the company on day $t$, denoted by 0 or 1. 

\begin{figure}[!h]
	\centering
	\subfigure[all customers]
	{
		\includegraphics[width=0.45\linewidth]{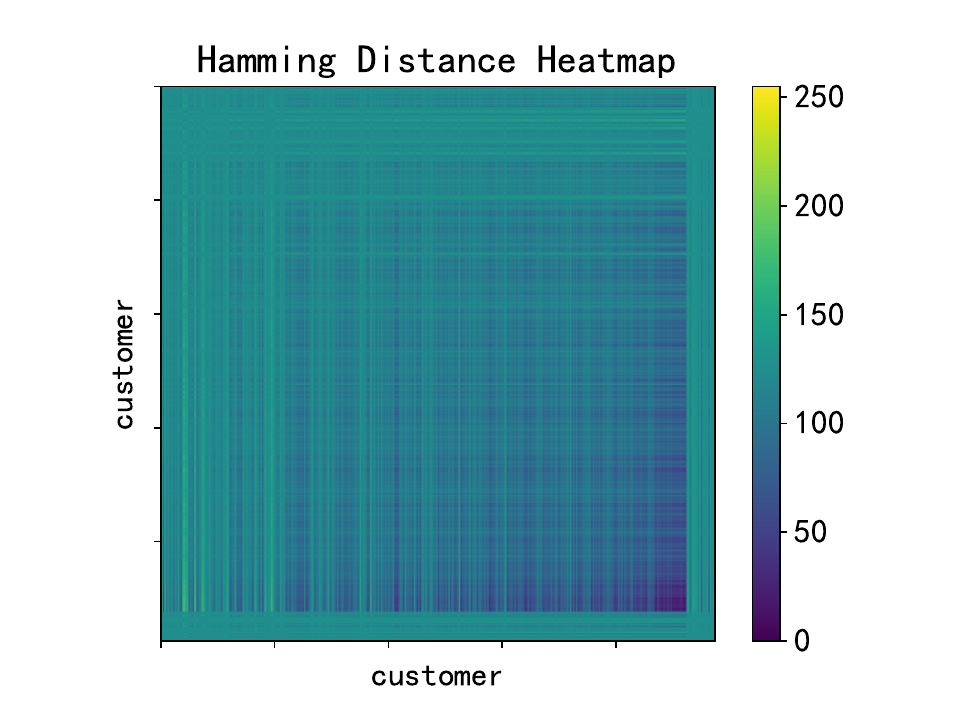}
		\label{fig:all customers}
	}
	%\hspace{0.01\linewidth} % ????
	\subfigure[cluster 1]
	{
		\includegraphics[width=0.45\linewidth]{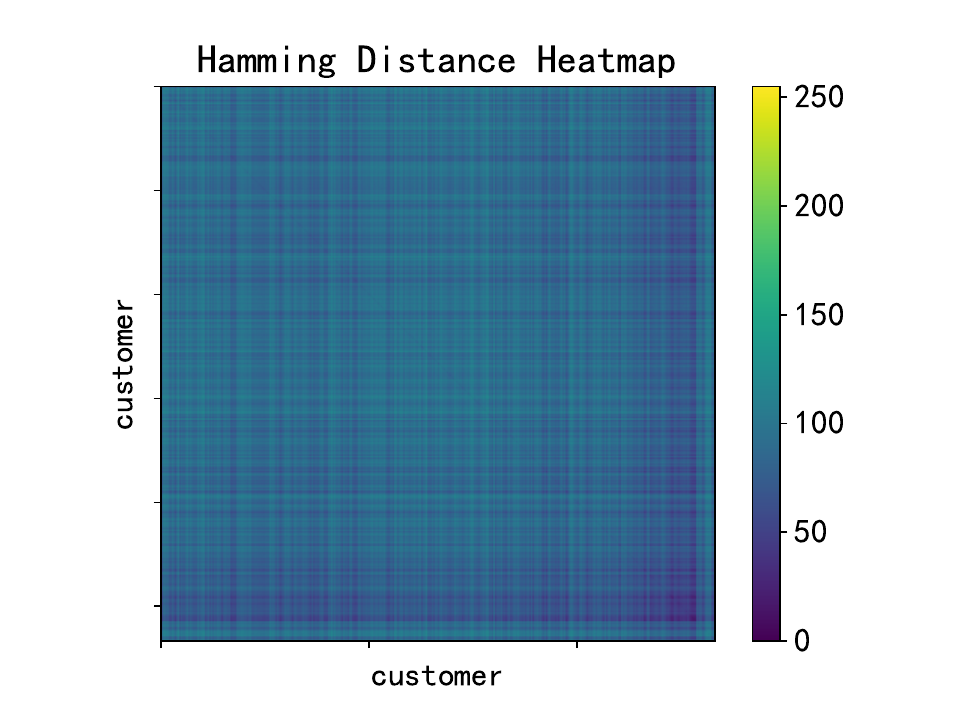}
		\label{fig:cluster 1}
	}
	%\hspace{0.01\linewidth} % ????
        \par
	\subfigure[cluster 2]
	{
		\includegraphics[width=0.45\linewidth]{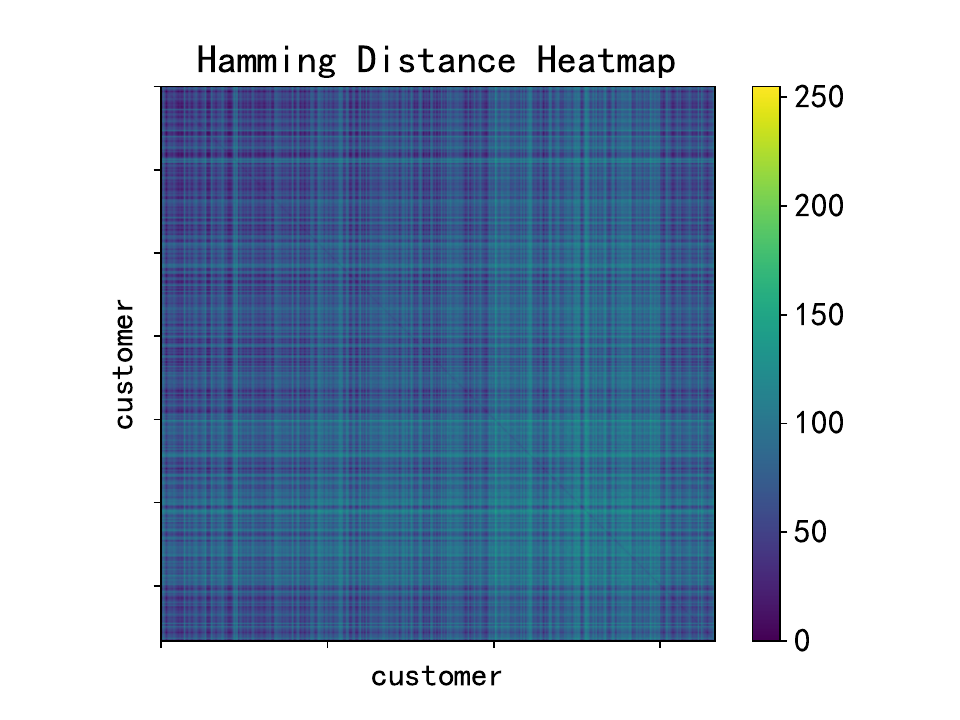}
		\label{fig:cluster 2}
	}
	%\hspace{0.07\linewidth} % ????	
	\subfigure[cluster 3]
	{
		\includegraphics[width=0.45\linewidth]{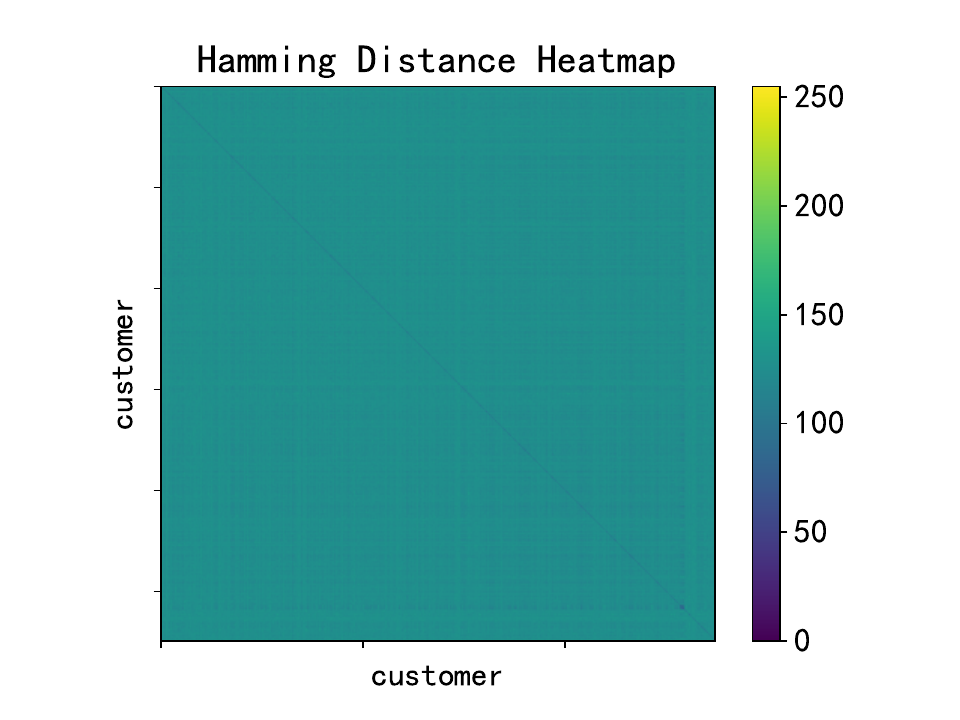}
		\label{fig:cluster 3}
	}	
	\caption{Comparison of the effects of the Hamming distance before and after clustering. The distribution of the Hamming distances before clustering is shown in~\ref{fig:all customers}, while~\ref{fig:cluster 1}-~\ref{fig:cluster 3} show the distributions of the three clusters after clustering.}
	\label{fig:Hamming_distance}
\end{figure}

\begin{figure}[h]
	\centering
	\subfigure[sample of all customers]
	{
		\includegraphics[width=0.45\linewidth]{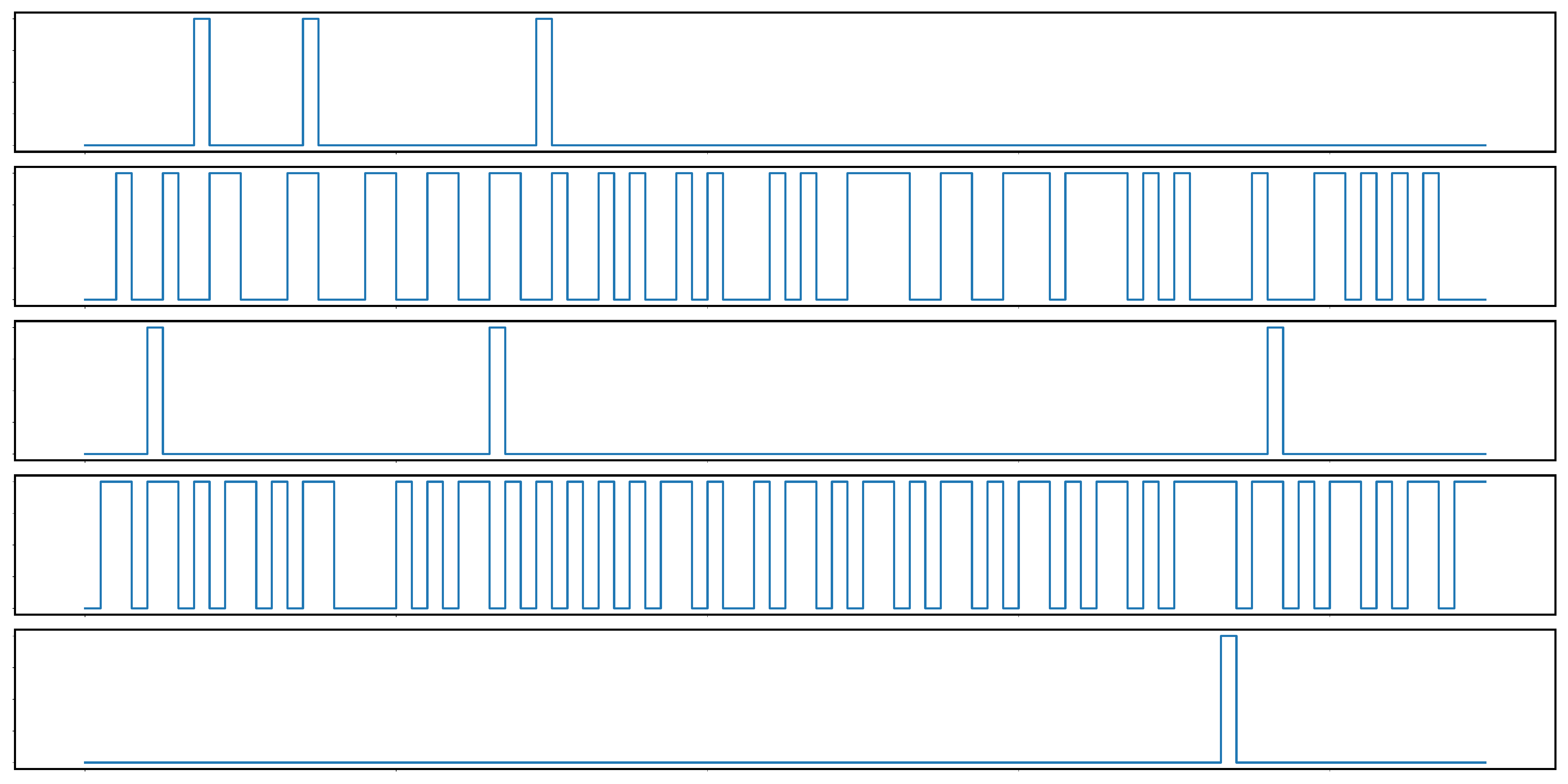}
		\label{fig:sample of all customers}
	}
	%\hspace{0.01\linewidth} % ????
	\subfigure[sample of cluster 1]
	{
		\includegraphics[width=0.45\linewidth]{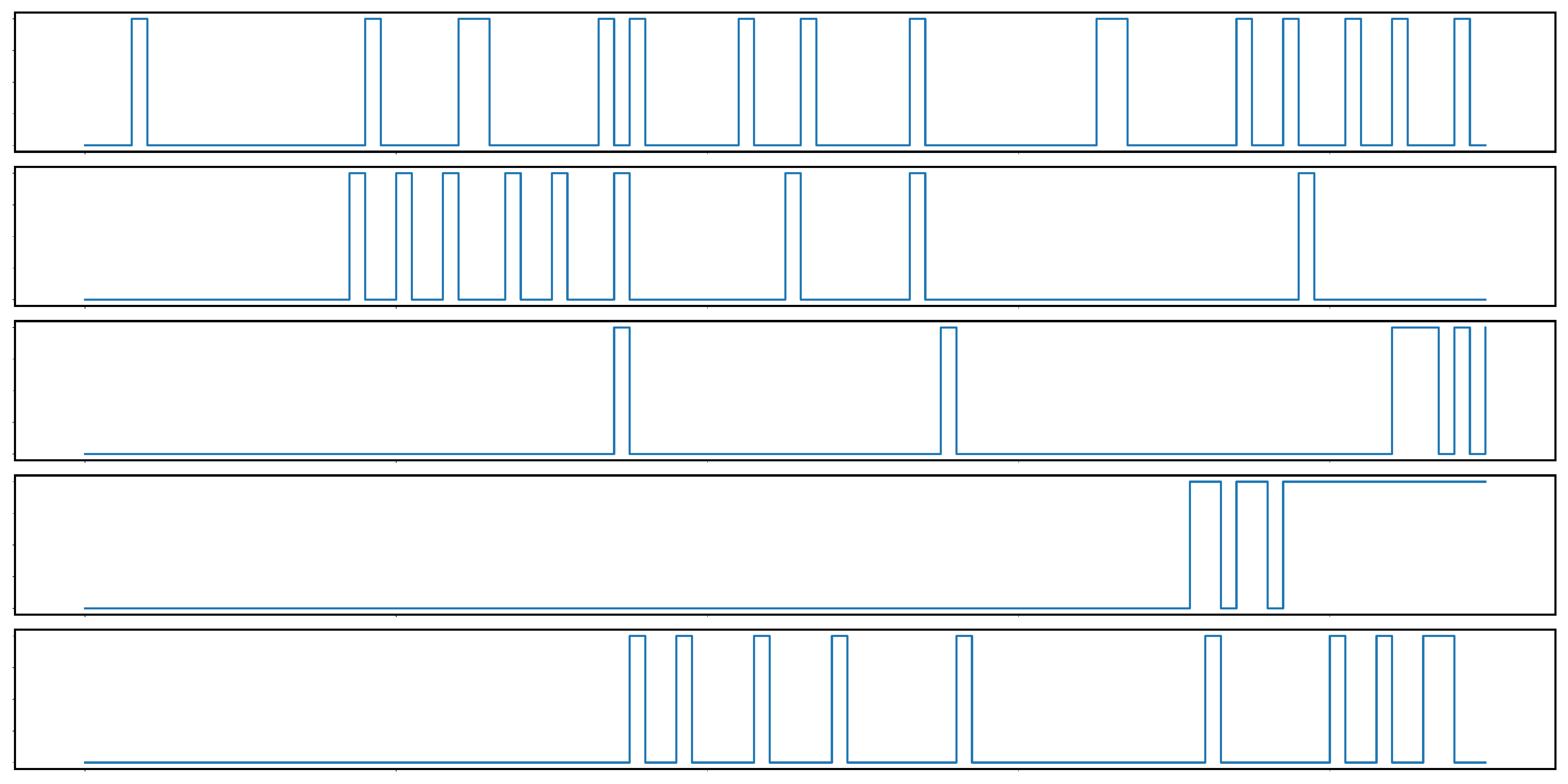}
		\label{fig:sample of cluster 1}
	}
	%\hspace{0.01\linewidth} % ????
        \par
	\subfigure[sample of cluster 2]
	{
		\includegraphics[width=0.45\linewidth]{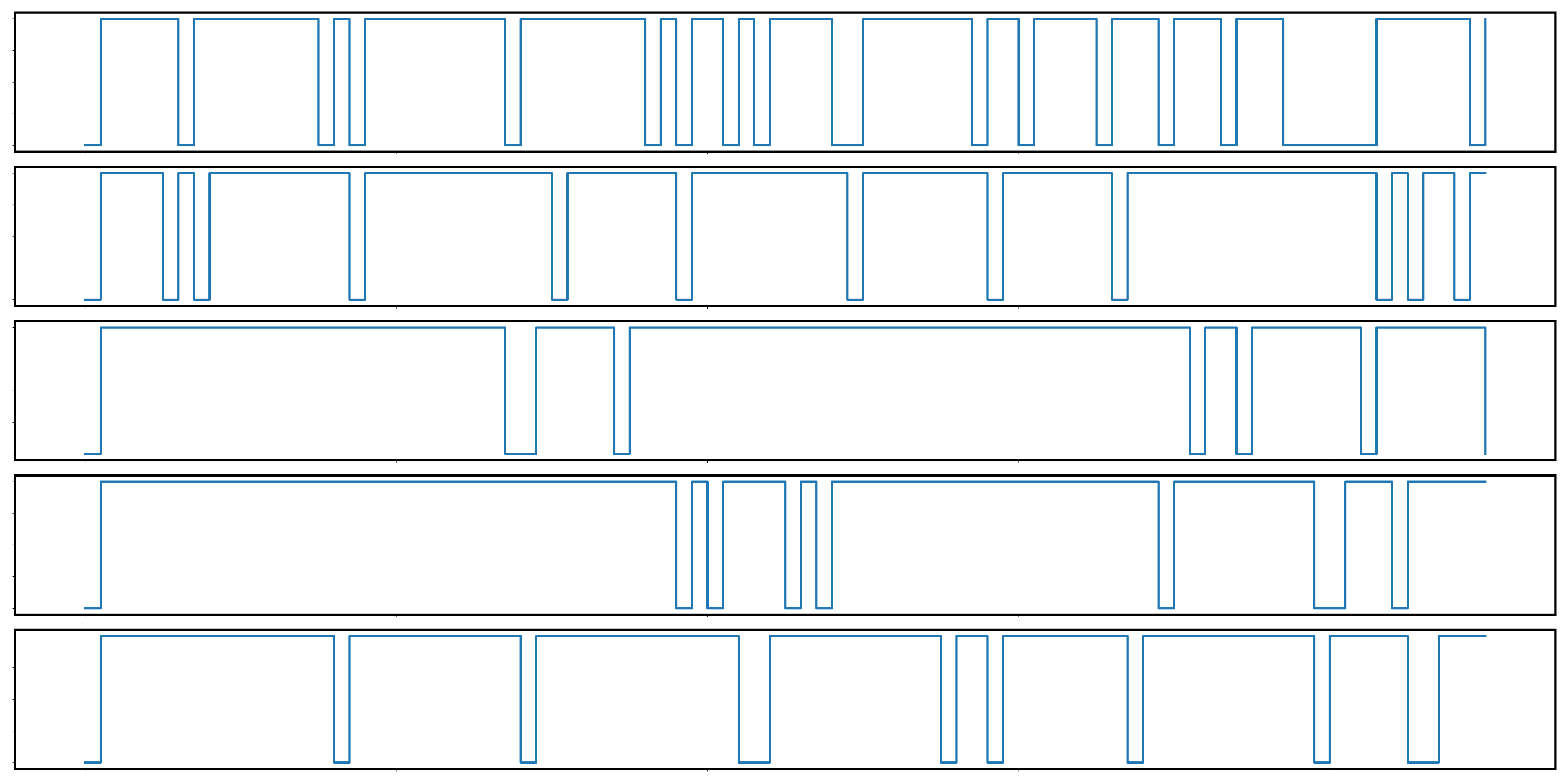}
		\label{fig:sample of cluster 2}
	}
	%\hspace{0.07\linewidth} % ????	
	\subfigure[sample of cluster 3]
	{
		\includegraphics[width=0.45\linewidth]{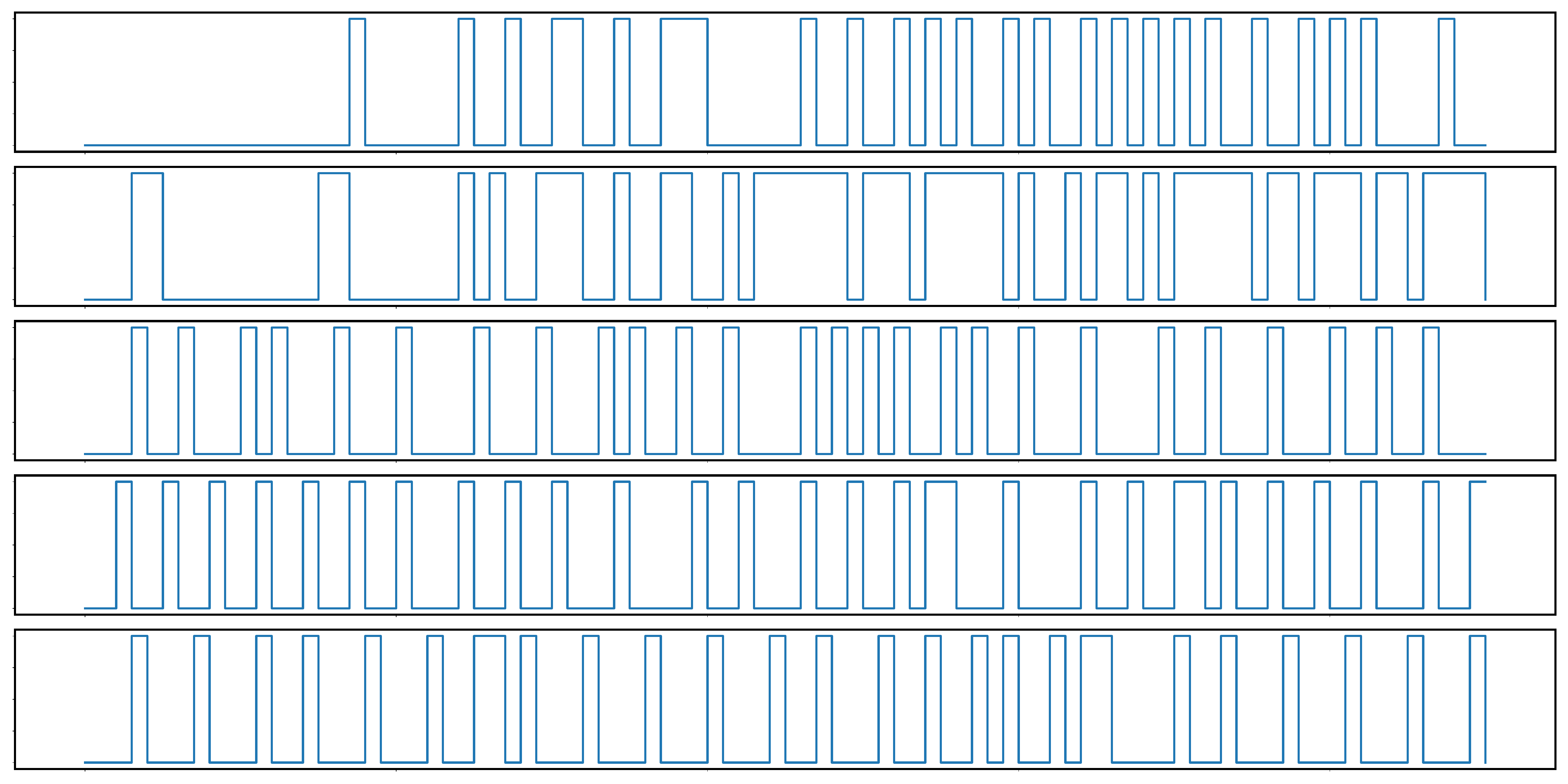}
		\label{fig:sample of cluster 3}
	}	
	\caption{The visualization of five randomly selected sample for each group.}
	\label{fig:sample of Hamming_distance}
\end{figure}

\textbf{Customer Hamming distance:} To measure the similarity between customer sequences, we subsequently extract the attendance sequences of 500 customers within this company and analyze the Hamming distance between these sequences. Given two customer sequences, $x_1$ and $x_2$, the Hamming distance (HD) is defined as follows:
\begin{equation}
	\text{HD}(x_1, x_2) = \sum_{i=1}^{n} \delta(x_1[i], x_2[i]),\quad x_1, x_2 \in X
\end{equation}
where $X$ denotes the set of all customer attendance sequences, $i$ denotes the position in the sequences, $n$ is the length of the sequences, and $\delta(\cdot)$ is an indicator function that takes a value of 1 when $a \neq b$ and is 0 otherwise. We compute the Hamming distance between all the sequences. As shown in \figurename~\ref{fig:all customers}, we give both a plot of the distribution results of the Hamming distance as well as five of the customers selected for sequence visualization in \figurename~\ref{fig:sample of all customers}. 

In the Hamming distance distribution plot, the horizontal and vertical coordinates represent the customers, and the color size represents the variability between two customers. For example, if the value is 50, it means that the variability of the two customers corresponding to the point is quantified as 50. Two conclusions can be drawn from the figures:

\begin{itemize}
    \item There is diversity among customers. Overall, customers exhibit a certain degree of diversity, indicating variability in their attendance patterns. This is characteristic of business marketing, where purchasing intentions among customers are often independent or have minimal influence on each other. Customers make purchases on the basis of fixed needs.
    \item There is similarity among customers. It is apparent that a considerable portion of customers exhibit low dissimilarity with other customers. Combined with \figurename~\ref{fig:CustActiveness}, these customers are likely to be either highly active or inactive. However, they may also belong to the moderately active customer group, where some customers share similar attendance patterns, displaying periodic attendance behaviors.	
\end{itemize}

To further investigate the similarity among these customers, we employ a clustering approach to uncover resemblances among these groups. Specifically, we conduct k-means clustering on the engagement levels of all the customers, resulting in $n$ clusters. Subsequently, we analyze the Hamming distance of attendance sequences corresponding to customers in different clusters. Here, $n$ is set to 3. As shown in \figurename~\ref{fig:cluster 1}-~\ref{fig:cluster 3}, we visualize the Hamming distance for the three clusters and give the results of sequence visualization for five customers selected from each cluster in Figure~\ref{fig:sample of cluster 1}-~\ref{fig:sample of cluster 3}. As can be seen from the figure, the similarity of the customers is further reflected in the four different groups. Overall, the more consistent the color, the smaller the difference, reflecting a higher degree of similarity between different customers within the group. When viewed in a separate column or row, the variability between customers is also significantly reduced. This is further evidence of the effectiveness of clustering.

\section{The Proposed Model}
\label{sec:Proposed Model}

\subsection{Overall Architecture}
\label{sec:Overall Architecture}
\begin{figure*}[h]
	\centering
	\centerline{
		\includegraphics[width=1.0\linewidth]{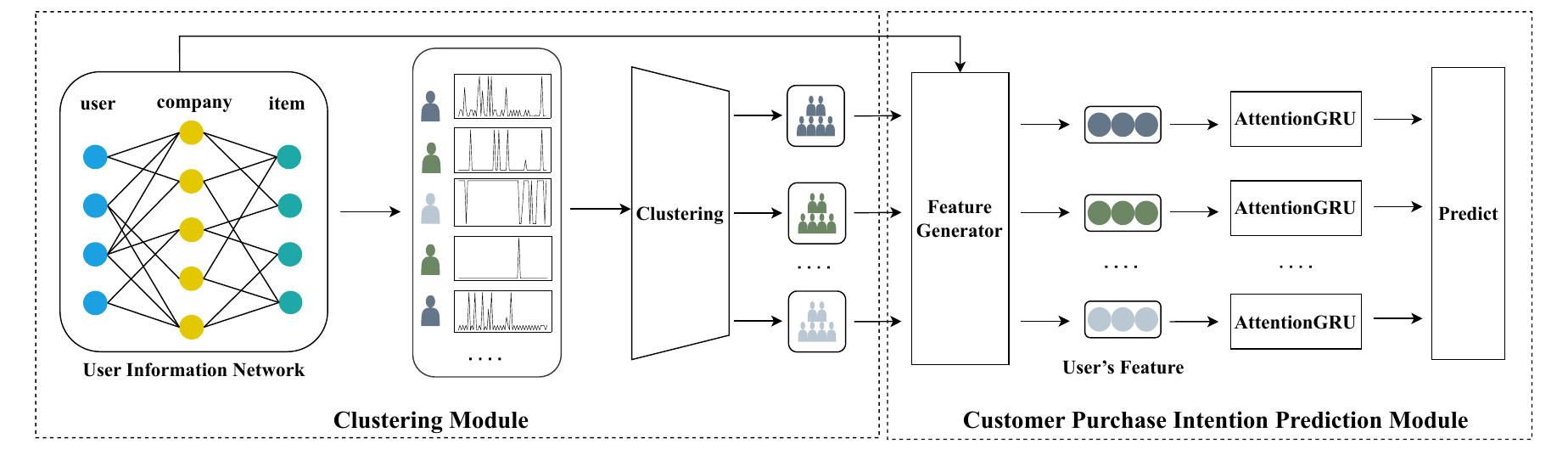}
	}
	\caption{Overview of the proposed CAGRU model}
	\label{fig:OverviewModel}
\end{figure*}

To effectively utilize customers' historical interaction data and to meet the business requirements of real-world scenarios, we propose a unified Clustering and Attention mechanism GRU model(CAGRU). As shown in~\figurename~\ref{fig:OverviewModel}, the framework consists of two main modules: a cluster module and a customer purchase intention prediction module. First, the clustering module is designed to aggregate similar customers and classify different customer groups, and then the customer purchase intention prediction module is designed to extract the time series characteristics of customer groups and make prediction. The details of the two modules are described below.

\subsection{Clustering Module}
\begin{figure}[!ht]
	\centering
	\centerline{
		\includegraphics[width=1.0\linewidth]{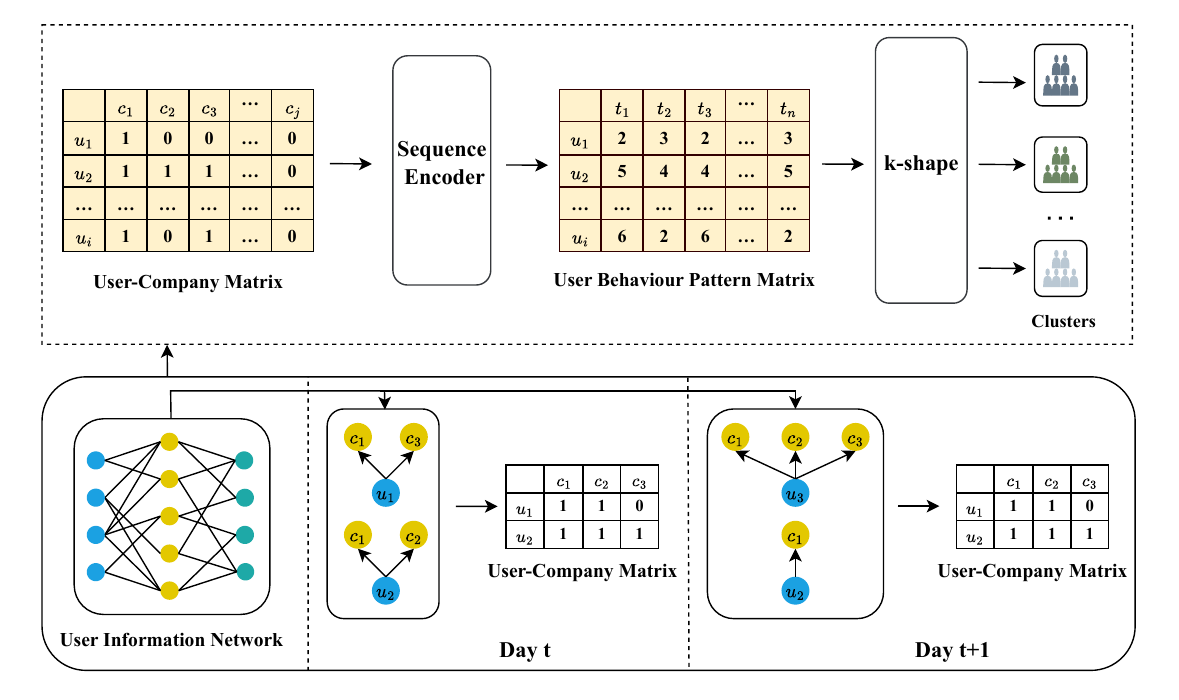}
	}
	\caption{Architecture diagram of the clustering module}
	\label{fig:ClusterModule}
\end{figure}

In this section, we introduce the clustering module we constructed, as shown in~\figurename~\ref{fig:ClusterModule}. The module consists of two components: sequence encoder and k-shape clustering. The details of the two components are described below.

\subsubsection{Sequence Encoder}
A user information network can be constructed on the basis of the transmission records of customers and companies. Customers generate different behavioral patterns based on different transactional behaviors, and we obtain an embedding of the customer's behavioral patterns by encoding them with a sequence encoder. We use $c_k$ to represent the $k$ shops under the company, and use $d_{u, c_k}^t \in \{0, 1\}$ to represent whether customer $u$ has made a purchase of shop $c_k$ on day $t$. Then, the purchases made by customer $u$ on day $t$ for $k$ shops under the company can be expressed as:
\begin{equation}
	N_t(u) = \{d_{u, c_1}^t, \dots, d_{u, c_k}^t\}
\end{equation}
where $N_t(u)$ denotes the customer's purchase interactions for each shop on day $t$ so that we can obtain the set of shop purchase behavior patterns for all customers for a total of $t$ days. Then a sequence encoder function is used to encode the behavioral patterns of the customers, and all the behavioral patterns constituted by the customer are jointly generated into a pattern dictionary $S$, denoted as follows:
\begin{equation}
	S_t(u_i)= g(N_t(u_i))
\end{equation}
where $S_t(u_i)$ represents the encoding mapping value corresponding to the sequence of purchasing behaviors generated by customer $u$ on day $t$. Eventually, we can obtain the sequence of patterns $X_n(u)$ for customer u on day n through the pattern set $S$, denoted as follows:
\begin{equation}
	X_n(u) = \{S_1(u), \dots, S_n(u)\}
\end{equation}
where $S_n(u)$ represents the buying behavioral pattern of customer $u$ corresponding to day $n$. 

\subsubsection{K-shape Clustering}
With the above encoder, we obtain a time series dataset containing the buying behavior patterns of all customers. Then we use the k-shape~\cite{paparrizos2015k} algorithm to cluster the time series, which is a distance metric based on shape-based distance (SBD) to capture the similarity of sequence shapes. For two sequences $x$ and $y$, their SBD distances are represented as follows:
\begin{equation}
	SBD(x, y)=1-\mathop {\max }\limits_m\frac{CC_w(x, y)}{ \sqrt{R_0(x, x)R_0(y, y)} }
\end{equation}
where $CC_w(x, y)$ denotes the covariance of $x$ and $y$ under weight $w$,  $R_0(x, x)$ and $R_0(y, y)$ denote the autocorrelation matrices of $x$ and $y$, respectively. The SBD obtained takes values ranging from 0 to 2, where 0 means that $x$ and $y$ are perfectly similar. 
Next, we compute the centre of mass according to the SBD algorithm. k-shape is used to find the centre of mass by computing the minimum of the sum of the squared distances between sequences, which will subsequently be divided to the centre of mass with the smallest distance to update the cluster membership. In this way, the clustered labels of the different customer groups are obtained. Based on the labels, we classify customer groups with similar characteristics, which are fed into the customer purchase intention prediction module for training and prediction. 

\subsection{Customer Purchase Intention Prediction Module}
\begin{figure}[h]
	\centering
	\centerline{
		\includegraphics[width=1.0\linewidth]{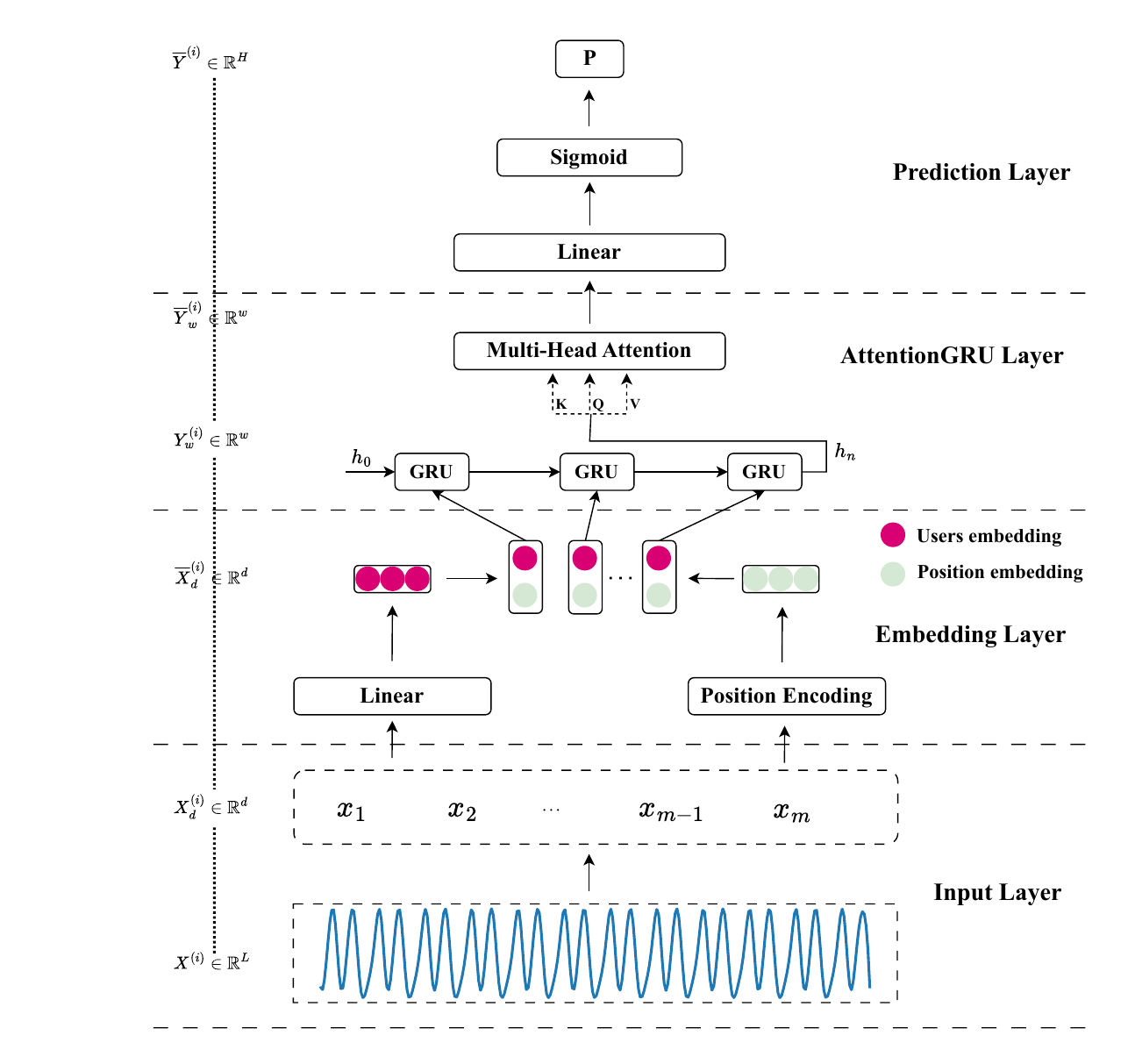}
	}
	\caption{Architecture diagram of the customer purchase intention prediction module.}
	\label{fig:AttentionGRUModule}
\end{figure}

The customer purchase intention prediction module is used to extract the time series features of different customer groups and predict them. As shown in \figurename~\ref{fig:AttentionGRUModule}, the module consists of four layers, including an input layer, an embedding layer, an attentionGRU layer and a prediction layer. The details are expressed as follows. 

\subsubsection{Input Layer and Embedding Layer}
The customer attributes include personal information, purchasing patterns, and recent transactions. For a customer, we aggregate its transaction sequence characteristics through a user information network, as described below:
\begin{equation}
	h_u = [p_u \oplus q_u]
\end{equation}
where $h_u$ represents the transactional features of the  customer, $p_u$ represents the personal attributes of the customer, and $q_u$ represents the attributes of the company, and $\oplus$  represents the aggregation operation. Preprocessing these features, we obtain the inputs $X^{(i)} \in \mathbb{R}^L$ to the model. 

\textbf{User Embedding.} Given a customer sequence channel $X^{(i)} \in \mathbb{R}^L$, sequences are obtained as user embedding $X_d^{(i)} \in \mathbb{R}^d$ through a layer of learnable linear projections, and d denotes the input dimension of the GRU. 

\textbf{Positional Embedding.} Practical marketing scenarios need to consider the timeliness of the customer's purchase, which day the customer arrives is often strongly associated with the recent trading activities. To this end, we introduce the relative position embedding method, which is calculated as follows:
\begin{equation}
	\text{PE}_{i, j} = 
	\begin{cases} 
		\sin\left(\frac{m}{{l}^{2j / d_{}}}\right) \quad n \in even\\
		\cos\left(\frac{m}{{l}^{2(n-1) / d_{}}}\right) \quad n \in odd
	\end{cases}
\end{equation}
where $l$ denotes the maximum length of the sequence, $m$ denotes the position index, and $n$ denotes the dimension index. We incorporate temporal weighting into the positional encoding of the sequence, and obtain the final sequence positional embedding $PE^{(i)} \in \mathbb{R}^d$ by multiplying the relative positional embedding of the sequence by a temporal decay factor, denoted as follows:
\begin{equation}
	\text{PE}_{i, j} = \alpha \text{PE}_{i, j} 
\end{equation}
where $\alpha$ denotes the temporal decay factor. Moreover, we concatenate the user embedding and positional embedding to represent the final sequence embedding $\overline X_d^{(i)} \in \mathbb{R}^d$, represented, denoted as follows:
\begin{equation}
	\overline X_d^{(i)} = concat(X_d^{(i)}, PE^{(i)})
\end{equation}
\subsubsection{AttentionGRU Layer and Prediction Layer}
We aggregate the user embedding and location embedding to obtain the inputs to the GRU. The GRU is a gate recursive unit based approach that employs a GRU to capture the temporal information of temporal data at different dates. The transformed $\overline X_d^{(i)}$ is fed into the GRU for loop iteration to capture temporal features. Specifically, for $x_t \in \mathbb{R}^d$ in $\overline X_d^{(i)}$, after $n$ iterations, the final hidden layer state feature $h_n$ can be obtained, which contains the historical features of the original sequence. We use $Y_w^{(i)} \in \mathbb{R}^w$ to denote the final output of the GRU, where $w$ denotes the output dimension of the GRU. The features of this part are passed to the attention mechanism module, which computes the relevance of the different elements in the input sequence through the attention mechanism and generates context vectors to enhance the model's understanding of the sequence. This module contains the linear transformation layer of the query, key and value. The attention weights are calculated via the $softmax$ function. The value vectors are weighted and summed according to the weights to obtain the context vectors. The output $\overline Y_w^{(i)} \in \mathbb{R}^w$ is obtained by the compression operation, and the computational process is represented as follows:
\begin{equation}
	\overline Y_w^{(i)} = softmax(q \times k^T) \times Y_w^{(i)}	
\end{equation}
where $q$, $k$ represent the query and key weight matrices, respectively.
Finally, $\overline Y_w^{(i)} \in \mathbb{R}^w$ goes through a linear projection layer and a sigmoid layer to obtain the final output probability $\overline Y^{(i)} \in \mathbb{R}^H$. 

\subsection{Loss calculation} 
Our task is to predict customer purchase intention, which is a binary classification task. Therefore, we choose Binary CrossEntropy(BCE) loss as the loss function for our model, defined as follows:
\begin{equation}
	\mathcal{L}_{\text{BCE}}(y, \hat{y}) = -\frac{1}{N} \sum_{i=1}^{N} \left[ y_i \log(\hat{y}_i) + (1 - y_i) \log(1 -               \hat{y}_i) \right]
\end{equation}
where $\mathcal{L}_{\text{BCE}}(y, \hat{y})$ denotes BCE loss, $y$ denotes the true label vector, $\hat{y}$ denotes the label vector predicted by the model, and $N$ denotes
the number of samples. $y_i$ and $\hat{y}_i$ denote the $i$ elements in the true label vector and the predicted label vector, respectively.

\section{Experiments}
\label{sec:Experiments}
This section initially outlines the experimental setup and performance metrics, followed by a comparison of the results with those of the state-of-the-art time series prediction models. Additionally, we conduct ablation studies to analyze the impact of clustering on the experimental performance. We also investigate the influence of crucial parameters on the CAGRU. All the experiments in this section are implemented in PyTorch and executed on a NVIDIA 1080Ti GPU with 11 GB of memory.

\subsection{Datasets and Evaluation Measures}
\subsubsection{Datasets}
As mentioned above, we conducted a study on a poultry company that mainly provides e-commerce services, and constructed different sizes of marketing datasets for this company according to the number of customer's company interactions, including CUSTOMER\_16K, CUSTOMER\_20K, CUSTOMER\_25K, and CUSTOMER\_30K. We divided each dataset into training, validation, and test sets based on the date length at a common ratio of 7:2:1. The validation set is used to tune the model parameters and the test set is used for performance comparison. The properties of the datasets are shown in Table~\ref{tab:tabel_dataset}. 
\begin{table}
	\caption{STATISTICS OF DATASET(K=1000). }
	\label{tab:tabel_dataset}
	\begin{tabular}{ccl}
		\toprule  % ????
		\textbf{Dataset} &  \textbf{\#Customer} &  \textbf{\#Interaction } \\
		\midrule  % ????
		CUSTOMER\_16K & 1.5K & 16K \\
		CUSTOMER\_20K & 1.6K & 20K \\
		CUSTOMER\_25K & 1.8K & 25K \\
		CUSTOMER\_30K & 2.0K & 30K \\
		\bottomrule % ????
	\end{tabular}
\end{table}
\subsubsection{Evaluation Measures}
Customer purchase intention prediction is essentially a binary classification task, where we predict whether a customer will make a purchase in the coming day. 
In this section we use common classification metrics to comprehensively measure the performance of our model, including accuracy (Acc), area under the curve (Auc), Precision, Recall, and F1Score. To fairly compare the metrics across the datasets, we take the top-$N$ of the prediction results to calculate the corresponding evaluation metrics. We define $N$ as 0.3 based on the average proportion of positive and negative samples in the dataset, i.e., the top 30\% of the probability of willingness to the output is set as 1, and the remaining values are set as 0. For model performance, a higher value of the five metrics represents a better performance.

\subsection{Comparison Experiments}
In this section we describe the baseline methods used for the comparison experiment and analyze for the validity of the CAGRU by comparing it with the existing baselines.

\begin{table*}[!ht]
	\huge
	\centering
	\caption{Comparison results on the four datasets. In each column, the best result is highlighted in bold and the suboptimal result is underlined. The improvement percentage(Imp) obtained by the CAGRU is defined as $ \frac{\text{The CAGRU result}-\text{The best baseline}}{\text{The best baseline}}$.}
	\resizebox{1\textwidth}{!}{
		\begin{tabular}{c|c|ccccccccc}
			\toprule
			\textbf{Dataset} & \textbf{Metric} & \textbf{LSTM} & \textbf{Transformer} & \textbf{LSTNet} & \textbf{Informer} & \textbf{TimesNet} & \textbf{Dlinear} & \textbf{PatchTST} & \textbf{CAGRU(Ours)} & \textbf{Imp} \\
			\midrule
			\multirow{5}[2]{*}{\textbf{CUSTOMER\_16K}} & \textbf{Acc} & \underline{0.7969} & 0.7209 & 0.764 & 0.7187 & 0.7425 & 0.7357 & 0.7459 & \textbf{0.8083} & 1.43\% \\
			& \textbf{Auc} & \underline{0.8473} & 0.714 & 0.786 & 0.7202 & 0.748 & 0.7458 & 0.7849 & \textbf{0.8566} & 1.10\% \\
			& \textbf{Precision} & \underline{0.5633} & 0.4367 & 0.5085 & 0.4329 & 0.4726 & 0.4612 & 0.4783 & \textbf{0.5822} & 3.36\% \\
			& \textbf{Recall} & \underline{0.7012} & 0.5435 & 0.6329 & 0.5388 & 0.5882 & 0.5741 & 0.5953 & \textbf{0.7247} & 3.35\% \\
			& \textbf{F1Score}& \underline{0.6247} & 0.4843 & 0.5639 & 0.4801 & 0.5241 & 0.5115 & 0.5304 & \textbf{0.6457} & 3.36\% \\
			\midrule
			\multirow{5}[2]{*}{\textbf{CUSTOMER\_20K}} & \textbf{Acc} & \underline{0.7961} & 0.6109 & 0.7427 & 0.6644 & 0.6805 & 0.6967 & 0.7154 & \textbf{0.8359} & 5.00\% \\
			& \textbf{Auc} & \underline{0.856} & 0.5579 & 0.7443 & 0.6406 & 0.6667 & 0.6817 & 0.7673 & \textbf{0.8803} & 2.84\% \\
			& \textbf{Precision} & \underline{0.6605} & 0.352 & 0.5714 & 0.441 & 0.4679 & 0.4948 & 0.5259 & \textbf{0.7267} & 10.02\% \\
			& \textbf{Recall} & \underline{0.6605} & 0.352 & 0.5714 & 0.441 & 0.4679 & 0.4948 & 0.5259 & \textbf{0.7267} & 10.02\% \\
			& \textbf{F1Score} & \underline{0.6605} & 0.352 & 0.5714 & 0.441 & 0.4679 & 0.4948 & 0.5259 & \textbf{0.7267} & 10.02\% \\
			\midrule
			\multirow{5}[2]{*}{\textbf{CUSTOMER\_25K}} & \textbf{Acc} & \underline{0.7811} & 0.6502 & 0.7668 & 0.6271 & 0.6304 & 0.692 & 0.7019 & \textbf{0.791} & 1.27\% \\
			& \textbf{Auc} & \underline{0.8351} & 0.5744 & 0.7884 & 0.5544 & 0.5466 & 0.6689 & 0.7293 & \textbf{0.8441} & 1.08\% \\
			& \textbf{Precision} & \underline{0.5275} & 0.3095 & 0.5037 & 0.2711 & 0.2766 & 0.3791 & 0.3956 & \textbf{0.544} & 3.13\% \\
			& \textbf{Recall} & \underline{0.6729} & 0.3949 & 0.6425 & 0.3458 & 0.3528 & 0.4836 & 0.5047 & \textbf{0.6939} & 3.12\% \\
			& \textbf{F1Score} & \underline{0.5914} & 0.347 & 0.5647 & 0.3039 & 0.3101 & 0.4251 & 0.4435 & \textbf{0.6099} & 3.13\% \\
			\midrule
			\multirow{5}[2]{*}{\textbf{CUSTOMER\_30K}} & \textbf{Acc} & \underline{0.7966} & 0.6334 & 0.764 & 0.6208 & 0.6547 & 0.705 & 0.6974 & \textbf{0.8418} & 5.67\% \\
			& \textbf{Auc} & \underline{0.8695} & 0.5789 & 0.8105 & 0.5842 & 0.6453 & 0.7069 & 0.7576 & \textbf{0.8905} & 2.42\% \\
			& \textbf{Precision} & \underline{0.6674} & 0.3954 & 0.613 & 0.3745 & 0.431 & 0.5146 & 0.5021 & \textbf{0.7426} & 11.27\% \\
			& \textbf{Recall} & \underline{0.6591} & 0.3905 & 0.6054 & 0.3698 & 0.4256 & 0.5083 & 0.4959 & \textbf{0.7338} & 11.33\% \\
			& \textbf{F1Score} & \underline{0.6632} & 0.3929 & 0.6091 & 0.3721 & 0.4283 & 0.5114 & 0.499 & \textbf{0.738} & 11.28\% \\
			\bottomrule
		\end{tabular}%
		\label{tab: Main_results}%
	}
\end{table*}%

\subsubsection{Baselines and Settings}
We choose seven well-known forecasting models as our baselines, including PatchTST~\cite{nie2022time}, LSTM~\cite{hochreiter1997long}, Informer~\cite{zhou2021informer}, Transformer~\cite{vaswani2017attention}, Dlinear~\cite{zeng2023transformers}, TimesNet~\cite{wu2022timesnet}, and LSTNet~\cite{lai2018modeling}.
In all the methods, we use the same embedding with the size set to 16. In our model, a different number of clustering clusters are used for each dataset for the best result, and the activation functions for the output and prediction layers are GeLU and Sigmoid, respectively. For the rest of the parameter settings, we follow the defaults suggested by the authors. 

\subsubsection{Comparison Results and Analysis}
The prediction results of the proposed CAGRU and other baselines on the four datasets are shown in Table~\ref{tab: Main_results}. We give the comparative results for the five categorical metrics, with the best, as well as the next best, scores as shown in bold and with underlined fonts, respectively. As seen from the table, the proposed CAGRU model outperforms all the baseline methods on all five metrics on these four datasets, which proves the validity of the model. Specifically, more significant improvements were achieved on the CUSTOMER\_30K dataset compared to the other three datasets. It can be seen that on the CUSTOMER\_30K dataset, compared to the next best baseline, the CAGRU method improves the Acc by 5.67\%, the Auc by 2.42\%, and the F1Score, by 11.28\%. This is because compared to the other three datasets, CUSTOMER\_30K dataset is richer in customer information and transaction records, and the clustering effect is more accurate. This suggests that the proposed CAGRU can effectively learn the similarity of customers, reduce the noise between sequences through clustering, and incorporate the attention mechanism to effectively learn the contextual information in the sequences, which can easily capture the short-term preferences of customers.

In addition, among all the methods, the LSTM achieves sub-optimal results on all four datasets. Because the nature of our task is short-term sequence prediction, the LSTM can take full advantage of capturing short-term customer preferences by using control gates without losing too much temporal information. Similarly, the proposed CAGRU method can maximize the retention of key information of the sequence through the gate mechanism, which proves that, compared with Transformers, CNNs, and MLPs, RNNs has certain advantages in short sequence prediction. Overall, the LSTNet method, which also incorporates LSTM and CNN, achieved the third best results among all the baselines, which further confirms that the RNN model has a strong advantage on our task. 

Furthermore, our comparison of the Transformer model and the Transformer-based models, namely, Informer and PatchTST, shows that Transformer performance is poor on all datasets, indicating that the initial Transformer prediction ability is weak. On the one hand, the Transformer architecture's dot product attention is insensitive to the local context, which may make the model prone to anomalies in the time series. On the other hand, the large size of the CUSTOMER dataset but the sparse nature of the data leads to a severe category imbalance problem, resulting in Transformer's inability to identify key information effectively. Similarly, Informer model underperforms because it focuses more on the long sequence prediction problem, owing to its generative decoder, which can effectively alleviate the context dependency problem of long sequences. However, it loses its advantage for short sequence prediction. Notably, PatchTST performs well on all four datasets because of its strategies of dividing the time series into patches and adopting channel independence, which can significantly reduce the computational complexity and retain key information.

\subsection{Ablation Studies}
\begin{figure}[h]
	\centering
	\subfigure[CUSTOMER\_16K]{
		\includegraphics[width=0.45\linewidth]{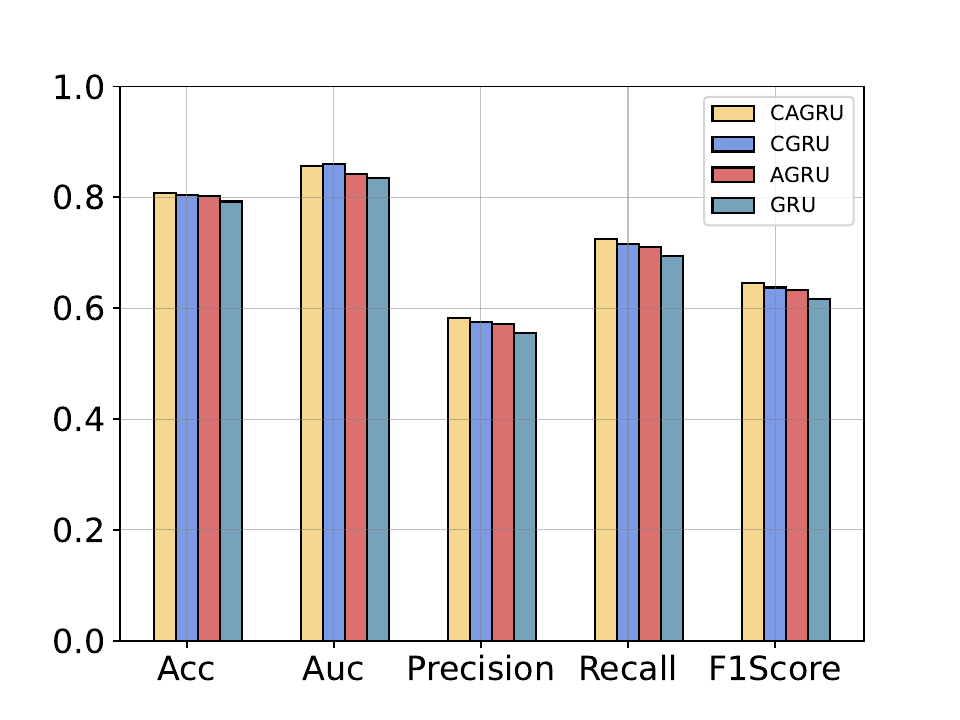}
		\label{fig:Ablation-dataset1}
	}
	%\hspace{0.02\linewidth} % ????
	\subfigure[CUSTOMER\_20K]{
		\includegraphics[width=0.45\linewidth]{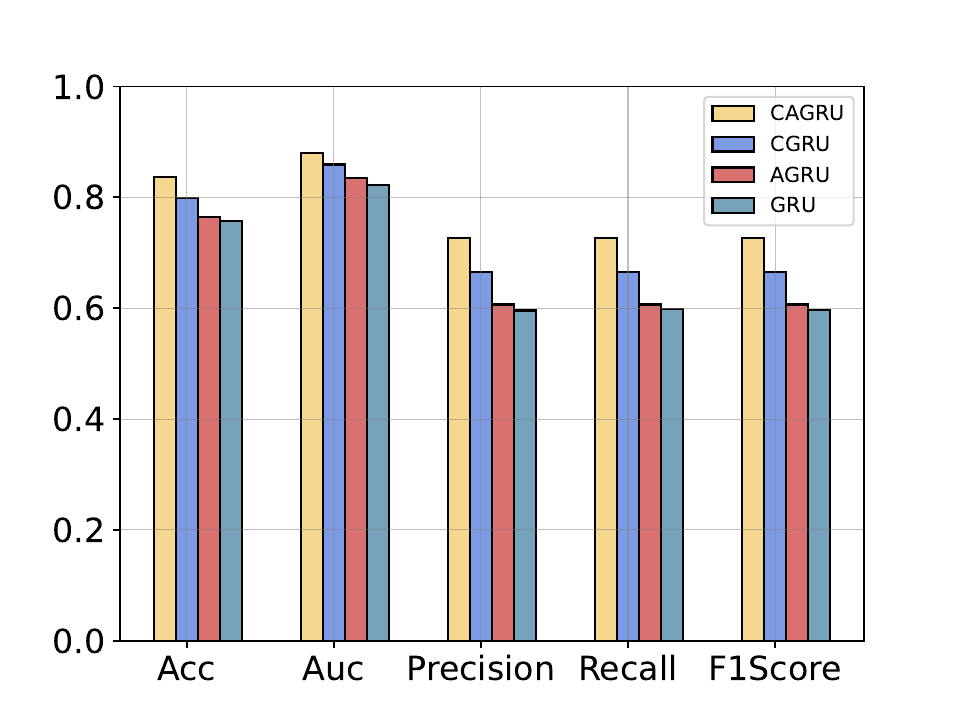}
		\label{fig:Ablation-dataset4}
	}
        \par
	%\hspace{0.03\linewidth} % ????
	\subfigure[CUSTOMER\_25K]{
		\includegraphics[width=0.45\linewidth]{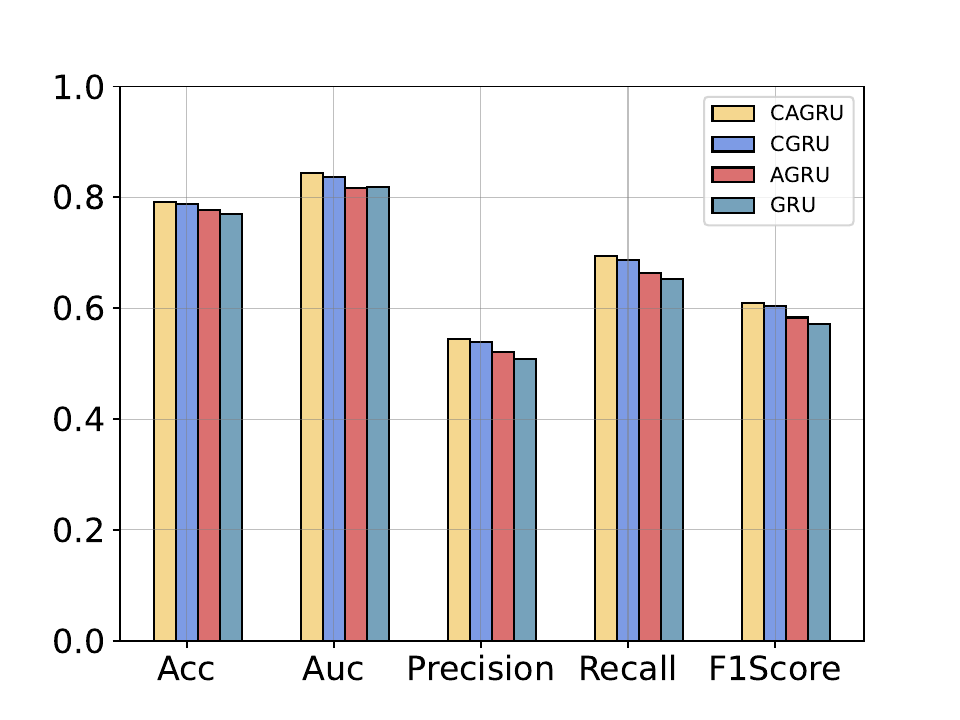}
		\label{fig:Ablation-dataset3}
	}
	%\hspace{0.02\linewidth} % ????
	\subfigure[CUSTOMER\_30K]{
		\includegraphics[width=0.45\linewidth]{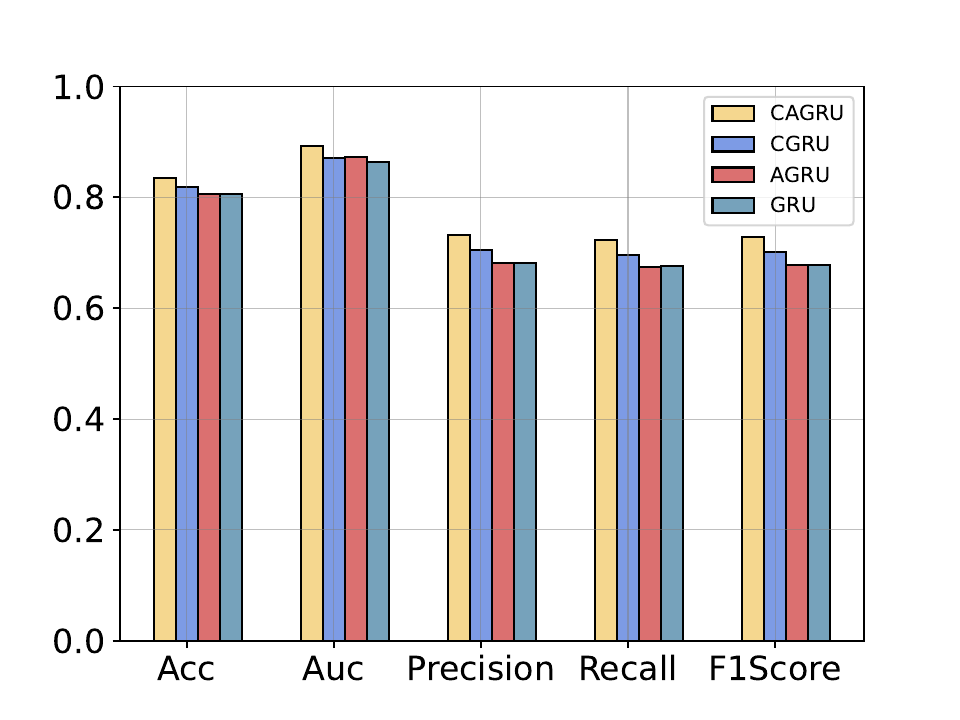}
		\label{fig:Ablation-dataset2}
	}
	\caption{Ablation study: Comparison results of the proposed CAGRU model with its three simplified variants on the four datasets.}
	\label{fig:AblationAnalysis}
\end{figure}

In this section, we explore the effects of the clustering module and the attention mechanism on model performance on each of the four datasets. To this end, we conduct three ablation studies that compare the proposed CAGRU with each of its three simplified variants. The first simplified variant, AGRU, removes the clustering module, which uses an attention mechanism to capture the similarity of customer sequences. The second variant is the CGRU, which removes the attention mechanism module and captures important contextual information by capturing key information about sequence positions. The third variant is the GRU, which retains only the original GRU structure, and captures the contextual dependencies in the sequences through the gate mechanism. The results are shown in \figurename~\ref{fig:AblationAnalysis}. According to the figure, the proposed CAGRU clearly outperforms the three simplified variants. The GRU is the worst, and the CGRU outperforms the AGRU. Moreover, a comparison of the CGRU and AGRU results reveals that the improvement of CGRU is more higher, which implies that adding a clustering module contributes more to model performance improvement than only considering the attention mechanism. The reason for this is that clustering captures the similarity of customer sequences and reduces the impact of customer plurality. Overall, the clustering module contributes more significantly to the proposed CAGRU method, which mitigates the effect of data sparsity on the model to a certain extent, and reduces the influence of groups with less feature information as well as the interaction information on other customer groups.

\begin{figure}[h]
	\centering
	\subfigure[CUSTOMER\_16K]
	{
		\includegraphics[width=0.45\linewidth]{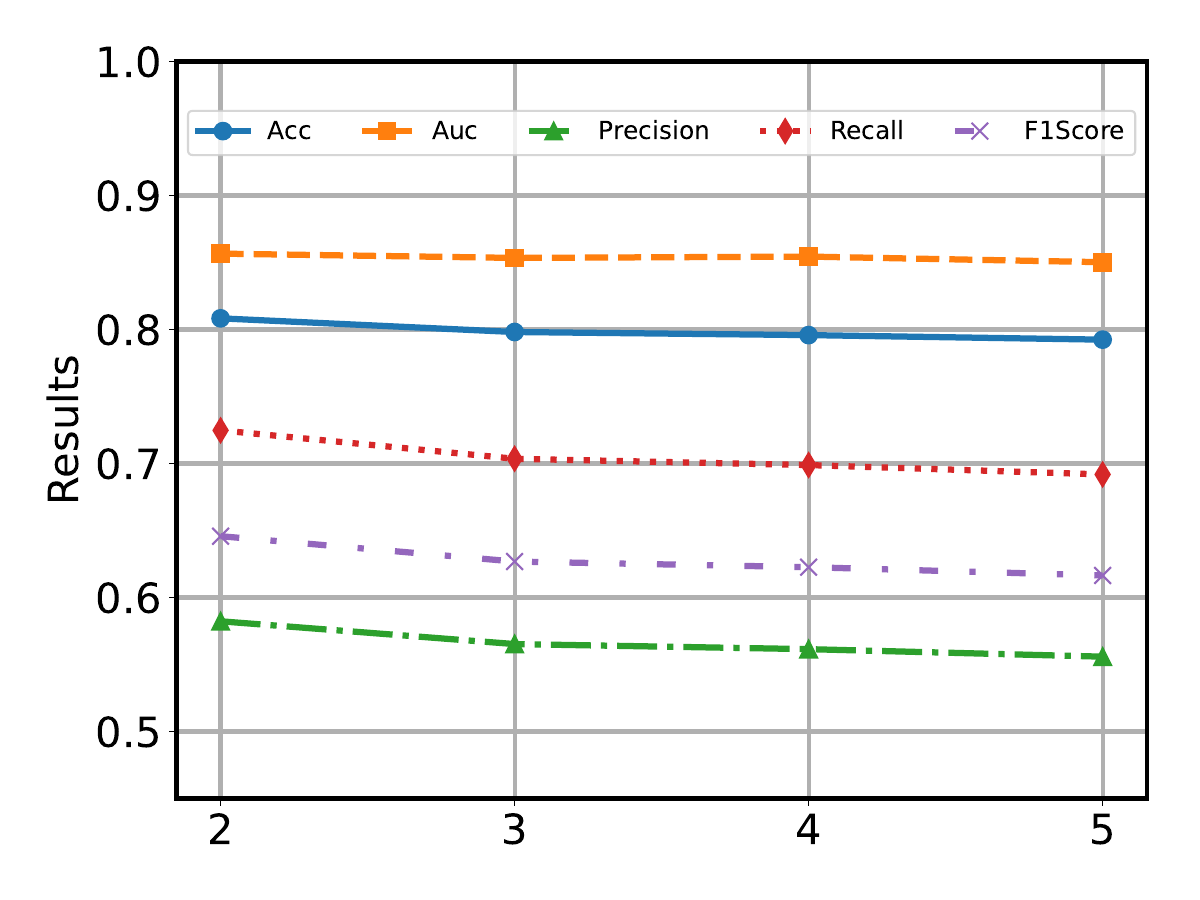}
		\label{fig:chao_clusters-dataset1}
	}
	%\hspace{0.02\linewidth} % ????
	\subfigure[CUSTOMER\_20K]
	{
		\includegraphics[width=0.45\linewidth]{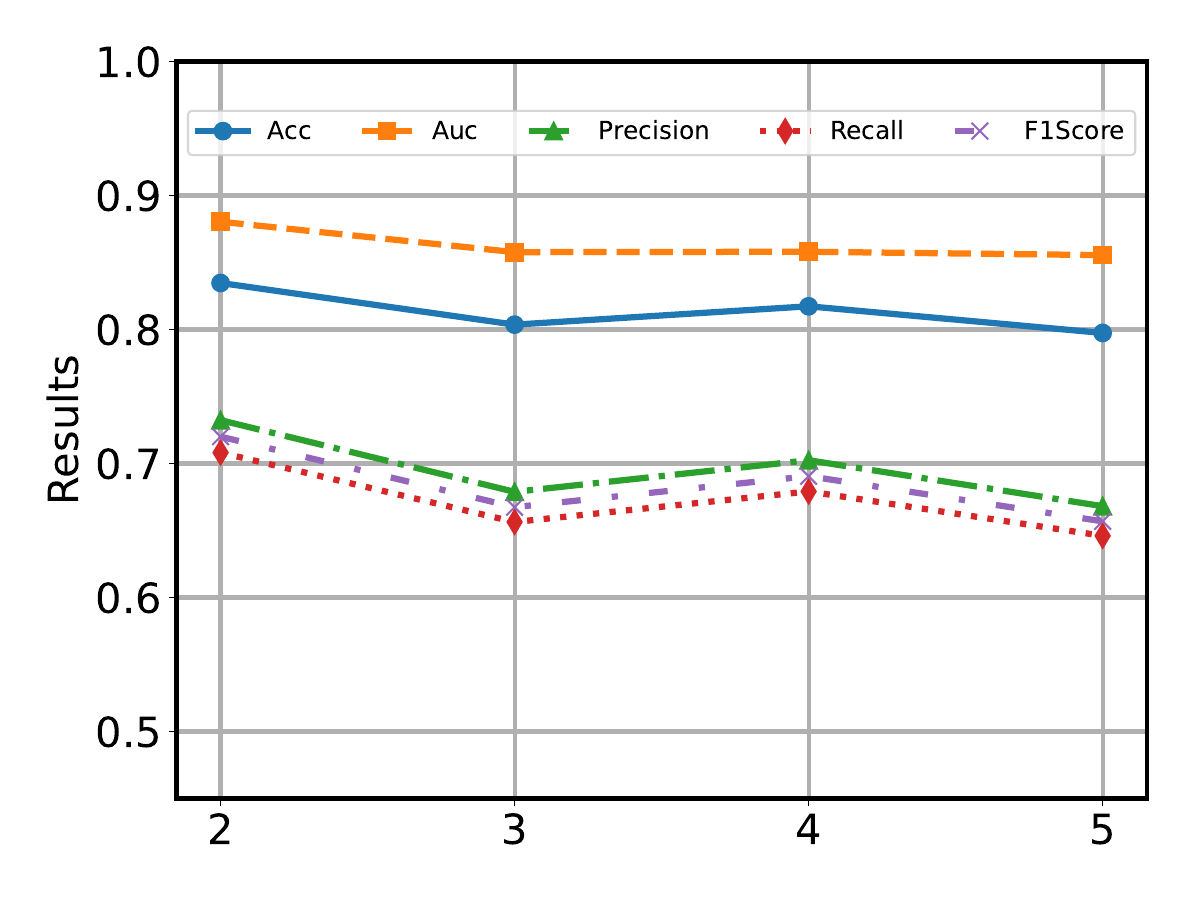}
		\label{fig:chao_clusters-dataset2}
	}
	%\hspace{0.02\linewidth} % ????
        \par
	\subfigure[CUSTOMER\_25K]
	{
		\includegraphics[width=0.45\linewidth]{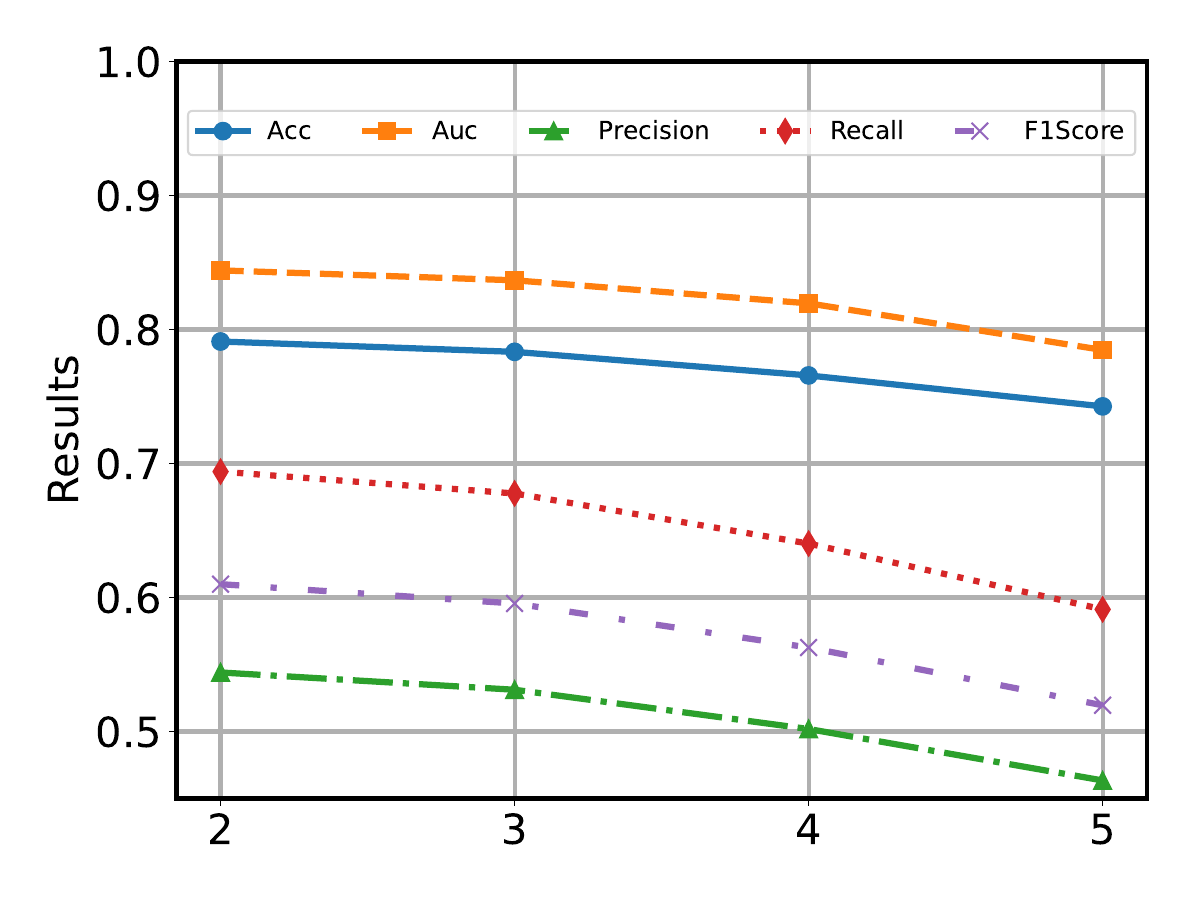}
		\label{fig:chao_clusters-dataset3}
	}
	%\hspace{0.02\linewidth} % ????	
	\subfigure[CUSTOMER\_30K]
	{
		\includegraphics[width=0.45\linewidth]{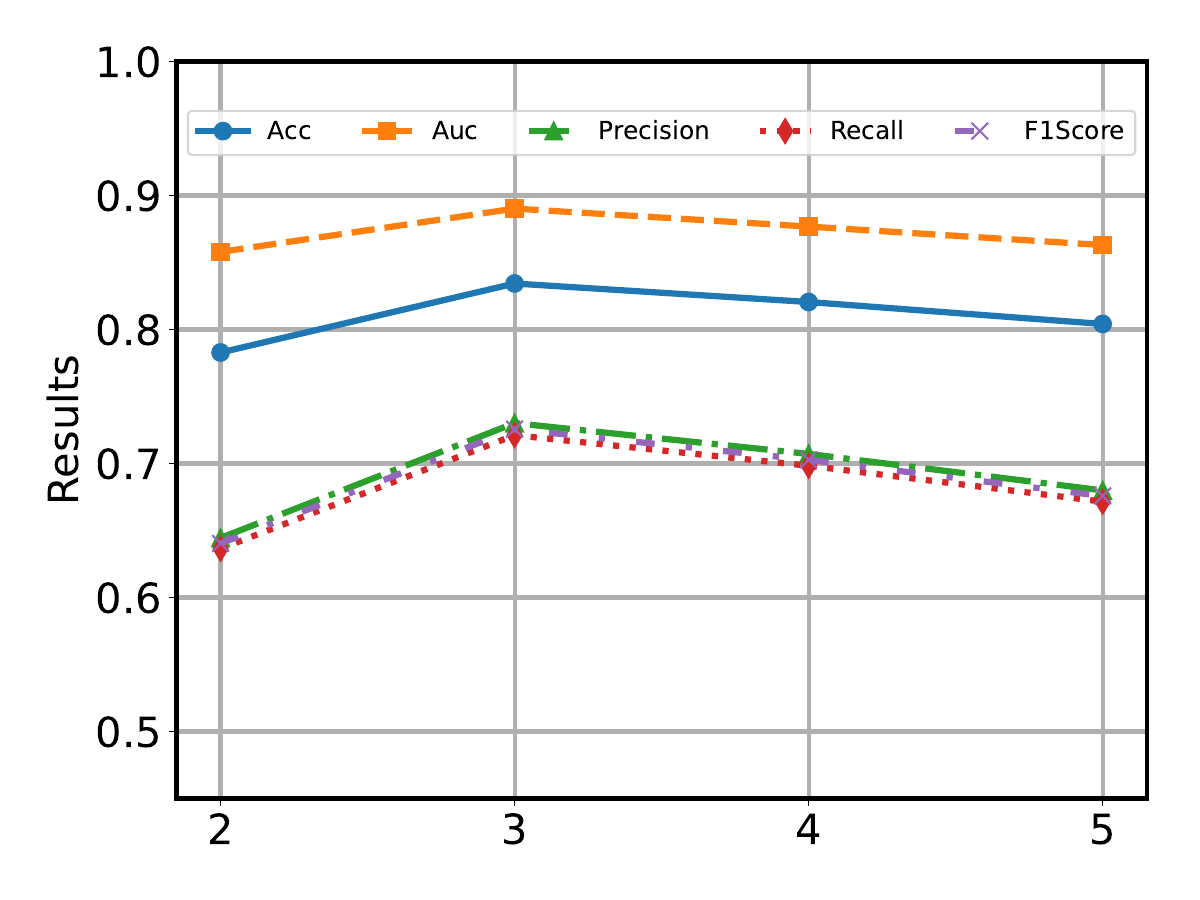}
		\label{fig:chao_clusters-dataset4}
	}	
	\caption{Parameter analysis: The impact of the number $n$ of 
  clusters on the performance of the proposed CAGRU on the four datasets. }
	\label{fig:ParameterAnalysis_cluster}
\end{figure}
\vspace{-\baselineskip} % 根据需要调整负值
\subsection{Parameter Analysis}
In this section, we analyze the effect of the number of clusters $n$ on CAGRU. We keep all other parameters constant and run CAGRU on the four datasets, varying the range of $n$ to $\{2, 3, 4, 5\}$. \figurename~\ref{fig:ParameterAnalysis_cluster} shows the impact of different cluster numbers on the metrics of each dataset. Except for CUSTPOMER\_30K, the other three datasets all achieve the best results at $n=2$, which basically matches the phenomenon of the long-tailed distribution of customer behaviors discussed in Section~\ref{sec:User_Survey}, where most of the customers belong to the long-term loyal customers or the occasional customers with low purchase frequency. The best results were achieved on the largest CUSTOMER\_30K on $n=3$, which shows that the clustering effect varies with the size of the customer. Overall, there is a fluctuating effect of different numbers of clusters on the performance of the proposed CAGRU model.

\section{Conclusions and Future Work}
\label{sec:Conclusions}
In this paper, we propose a clustering-based time series prediction algorithm for customer purchase intention called CAGRU. Aiming at the phenomenon that customer groups show head-to-tail distribution, we explore the similarities and differences among customers to classify them into different groups via a clustering algorithm. Then, incorporating multi-modal data, we input them into GRU network incorporating the attention mechanism to effectively capture the sequential features of customers' purchasing behaviors, facilitating the prediction accuracy of customer purchase intention. We conducted extensive experiments on four e-commerce datasets to validate the superiority and effectiveness of the proposed CAGRU model. Future research could explore long sequence customer purchase intention prediction. It is also interesting to extend the proposed method to other scenarios, where scenarios involving multivariate time series and multi-modal data can also make use of time series clustering to reduce the instability of the downstream task in capturing the features among the multivariate variables.

%%
%% The acknowledgments section is defined using the "acks" environment
%% (and NOT an unnumbered section). This ensures the proper
%% identification of the section in the article metadata, and the
%% consistent spelling of the heading.
\begin{acks}
This research was funded in part by National Key Research and Development Program of China (2023YFD1301903)
\end{acks}

%%
%% The next two lines define the bibliography style to be used, and
%% the bibliography file.
\bibliographystyle{ACM-Reference-Format}
\balance
\bibliography{sample-base}

%%
%% If your work has an appendix, this is the place to put it.
\appendix

\end{document}